# Magnonic Holographic Memory: from Proposal to Device

F. Gertz[1], A. Kozhevnikov[2], Y. Filimonov[2], D.E. Nikonov[3] and A. Khitun[1]

[1])Electrical Engineering Department, University of California - Riverside, Riverside, CA, USA, 92521
[2])Kotel'nikov Institute of Radioengineering and Electronics of Russian Academy of Sciences, Saratov Branch, Saratov, Russia, 410019
[3])Technology & Manufacturing Group Intel Corp. , 2501 NW 229th Avenue, Hillsboro, OR, USA, 97124

In this work, we present recent developments in magnonic holographic memory devices exploiting spin waves for information transfer. The devices comprise a magnetic matrix and spin wave generating/detecting elements placed on the edges of the waveguides. The matrix consists of a grid of magnetic waveguides connected via cross junctions. Magnetic memory elements are incorporated within the junction while the read-in and read-out is accomplished by the spin waves propagating through the waveguides. We present experimental data on spin wave propagation through NiFe and YIG magnetic crosses. The obtained experimental data show prominent spin wave signal modulation (up to 20 dB for NiFe and 35 dB for YIG) by the external magnetic field, where both the strength and the direction of the magnetic field define the transport between the cross arms. We also present experimental data on the 2-bit magnonic holographic memory built on the double cross YIG structure with micro-magnets placed on the top of each cross. It appears possible to recognize the state of each magnet via the interference pattern produced by the spin waves with all experiments done at room temperature. Magnonic holographic devices aim to combine the advantages of magnetic data storage with wave-based information transfer. We present estimates on the spin wave holographic devices performance, including power consumption and functional throughput. According to the estimates, magnonic holographic devices may provide data processing rates higher than $1\times10^{18}$ bits/cm$^2$/s while consuming 0.15mW. Technological challenges and fundamental physical limits of this approach are also discussed.

## I. INTRODUCTION

THERE is growing interest in novel computational devices able to overcome the limits of the current complimentary-metal-oxide-semiconductor (CMOS) technology and provide further increase of the computational throughput [1]. So far, the majority of the "beyond CMOS" proposals are aimed at the development of new switching technologies[2, 3] with increased scalability and improved power consumption characteristics over the silicon transistor. However, it is difficult to expect that a new switch will outperform CMOS in all figures of merit, and more importantly, will be able to provide multiple generations of improvement as was the case for CMOS [4]. An alternative route to the computational power enhancement is via the development of novel computing devices *aimed not to replace* but *to complement CMOS* by special task data processing [5]. Spin wave (magnonic) logic devices are one of the alternative approaches aimed to take the advantages of the wave interference at nanometer scale and utilize phase in addition to amplitude for building logic units for parallel data processing.

A spin wave is a collective oscillation of spins in a magnetic lattice, analogous to phonons, the collective oscillation of the nuclear lattice. The typical propagation speed of spin waves does not exceed $10^7$cm/s, while the attenuation time at room temperature is about a nanosecond in the conducting ferromagnetic materials (e.g. NiFe, CoFe) and may be hundreds of nanoseconds in non-conducting materials (e.g. YIG). Such a short attenuation time explains the lack of interest in spin waves as a potential information carrier in the past. The situation has changed drastically as the technology of integrated logic circuits has scaled down to the deep sub-micrometer scale, where the short propagation distance of spin waves (e.g. tens of microns at room temperature) is more than sufficient for building logic circuits. At the same time, spin waves have several inherent appealing properties making them promising for building wave-based logic devices. For instance, spin wave propagation can be directed by using magnetic waveguides similar to optical waveguides. The amplitude and the phase of propagating spin waves can be modulated by an external magnetic field. Spin waves can be generated and detected by electronic components (e.g. multiferroics [6]), which make them suitable for integration with conventional logic circuits. Finally, the coherence length of spin waves at room temperature may exceed tens of microns, which allows for the utilization of spin wave interference for logic functionality. It makes spin waves much more prone to scattering than a single electron and resolves one of the most difficult problems of spintronics associated with the necessity to preserve spin orientation while transmitting information between the spin-based units.

During the past decade, there have been a growing number of theoretical and experimental works exploring spin wave propagation in a variety of magnetic structures [7, 8], the possibility of spin wave propagation modulation by an external magnetic field [9, 10], and spin wave interference and diffraction [11-15]. The collected experimental data revealed interesting and unique properties of spin wave transport (e.g. non-reciprocal spin wave propagation [16]) for building magnonic logic circuits. The first working spin-wave based logic device has been experimentally demonstrated by Kostylev *et al* in 2005[17]. The authors used the Mach–Zehnder-type current-controlled spin wave interferometer to demonstrate output voltage modulation as a result of spin wave interference. Later on, exclusive-not-OR (NOR) and not-AND (NAND) gates were experimentally demonstrated utilizing a similar structure [18]. The idea of using Mach–Zehnder-type spin wave interferometers has been further evolved by proposing a spin wave interferometer with a vertical current-carrying wire [19]. With zero applied current, the spin waves in two branches interfere constructively and propagate through the structure. The waves interfere destructively and do not propagate through the structure if a certain electric current is applied. At some point, these first magnonic logic devices resemble the classical field effect transistor, where the magnetic field produced by the electric current modulates the propagation of the spin wave—an analogue to the electric current. Then, it was proposed to combine spin wave with nano-magnetic logic aimed to combine the advantages of non-volatile data storage in magnetic memory and the enhanced functionality provided by the spin wave buses [20]. The use of spin wave interference makes it possible to realize Majority gates (which can be used as AND or OR gates) and NOT gates with a fewer number of elements than is required for transistor-based circuitry, promising the further reduction of the size of the logic gates. There were several experimental works demonstrating three-input spin wave Majority gates [21,22]. However, the integration of the spin wave buses with nano-magnets in a digital circuit, where the magnetization state of the nano-magnet is controlled by a spin wave has not yet been realized.

An alternative approach to spin wave-based logic devices is to build non-Boolean logic gates for special task data processing. The essence of this approach is to maximize the advantage of spin wave interference. Wave-based analog logic circuits are potentially promising for solving problems requiring parallel operation on a number of bits at time (i.e. image processing, image recognition). The concept of magnonic holographic memory (MHM) for data storage and special task data processing has been recently proposed [23]. Holographic devices for data processing have been extensively developed in optics during the past five decades. The development of spin wave-based devices allows us to implement some of the concepts developed for optical computing to magnetic nanostructures utilizing spin waves instead of optical beams. There are certain technological advantages that make the spin wave approach even more promising than optical computing. First, short operating wavelength (i.e. 100nm and below) of spin wave devices promises a significant increase of the data storage density ($\sim\lambda^2$ for 2D and $\sim\lambda^3$ for 3D memory matrixes). Second, even more importantly, is that spin wave bases devices can have voltage as an input and voltage as an output, which makes them compatible with the conventional CMOS circuitry. Though spin waves are much slower than photons, magnonic holographic devices may possess a higher memory capacity due to the shorter operational wavelength and can be more suitable for integration with the conventional electronic circuits. In this work, we present recent experimental results on magnonic holographic memory and discuss the advantages and potential shortcomings of this approach. The rest of the paper is organized as follows. In Section II, we describe the structure and the principle of operation of magnonic holographic memory. Next, we present experimental data on the first 2-bit magnonic holographic memory in Section III. The advantages and the challenges of the magnonic holographic devices are discussed in the Sections IV. In Section V, we present the estimates on the practically achievable performance characteristics.

II. MATERIAL STRUCTURE AND THE PRINCIPLE OF OPERATION

The schematics of a MHM device are shown in Figure 1(A). The core of the structure is a magnetic matrix consisting of the grid of magnetic waveguides with nano-magnets placed on top of the waveguide junctions. Without loss of generality, we have depicted a 2D mesh of orthogonal magnetic waveguides, though the matrix may be realized as a 3D structure comprising the layers of magnetic waveguides of a different topology (e.g. honeycomb magnetic lattice). The waveguides serve as a media for spin wave propagation – spin wave buses. The buses can be made of a magnetic material such as yttrium iron garnet $Y_3Fe_2(FeO_4)_3$ (YIG) or permalloy ($Ni_{81}Fe_{19}$) ensuring maximum possible group velocity and minimum attenuation for the propagating spin waves at room temperature. The nano-magnets placed on top of the waveguide junctions act as memory elements holding information encoded in the magnetization state. The nano-magnet can be designed to have two or several thermally stable states of magnetization, where the number of states defines the number of logic bits stored in each junction. The spins of the nano-magnet are coupled to the spins of the junction magnetic wires via the exchange and/or dipole-dipole coupling affecting the phase of the propagation of spin waves. The phase change received by the spin wave depends on the strength and direction of the magnetic field produced by the nano-magnet. At the same time, the spins of the nano-magnet are affected by the local magnetization change caused by the propagating spin waves.

The input/output ports are located at the edges of the waveguides. These elements are aimed to convert the input electric signals into spin waves, and vice versa, convert the output spin waves into electrical signals. There are several possible options for building such elements by using micro-antennas [24, 25], spin torque oscillators[26], and multiferroic elements[6]. For example, the micro-antenna is a conducting contour placed in the vicinity of the spin wave bus. An electric current passed through the contour generates a magnetic field around the current-carrying wires, which excites spin waves in the magnetic material, and vice versa, a propagating spin wave changes the magnetic flux from the magnetic waveguide and generates an inductive voltage in the antenna contour. The advantages and shortcomings of different input/output elements will be discussed later in the text.

Spin waves generated by the edge elements are used for information read-in and read-out. The difference among these two modes of operation is in the amplitude of the generated spin waves. In the read-in mode, the elements generate spin waves of a relatively large amplitude, so two or several spin waves coming in-phase to a certain junction produce magnetic field sufficient for magnetization change within the nano-magnet. In the read-out mode, the amplitude of the generated spin waves is much lower than the threshold value required to overcome the energy barrier between the states of nano-magnets. So, the magnetization of the junction remains constant in the read-out mode. The details of the read-in and read-out processes are presented in Ref. [23].

The formation of the hologram occurs in the following way. The incident spin wave beam is produced by the number of spin wave generating elements (e.g. by the elements on the left side of the matrix as illustrated in Figure 1(B)). All the elements are biased by the same RF generator exciting spin waves of the same frequency, $f$, and amplitude, $A_0$, while the phase of the generated waves are controlled by DC voltages applied individually to each element. Thus, the elements constitute a phased array allowing us to artificially change the angle of illumination by providing a phase shift between the input waves. Propagating through the junction, spin waves accumulate an additional phase shift, $\Delta\phi$, which depends on the strength and the direction of the local magnetic field provided by the nano-magnet, $H_m$:

$$\Delta\phi = \int_0^r k(\vec{H}_m)dr , \quad (1)$$

where the particular form of the wavenumber $k(H)$ dependence varies for magnetic materials, film dimensions, the mutual direction of wave propagation and the external magnetic field[27]. For example, spin waves propagating perpendicular to the external magnetic field (magnetostatic surface spin wave – MSSW) and spin waves propagating parallel to the direction of the external field (backward volume magnetostatic spin wave – BVMSW) may obtain significantly different phase shifts for the same field strength. The phase shift $\Delta\phi$ produced by the external magnetic field variation $\delta H$ in the ferromagnetic film can be expressed as follows[17]:

$$\frac{\Delta\phi}{\partial H} = \frac{l}{d}\frac{(\gamma H)^2 + \omega^2}{2\pi\gamma^2 M_s H^2} \quad \text{(BVMSW)},$$

$$\frac{\Delta\phi}{\delta H} = -\frac{l}{d}\frac{\gamma^2(H + 2\pi M_s)}{\omega^2 - \gamma^2 H(H + 4\pi M_s)} \quad \text{(MSSW)}, \quad (2)$$

where $\Delta\phi$ is the phase shift produced by the change of the external magnetic field $\delta H$, $l$ is the propagation length, $d$ is the thickness of the ferromagnetic film, $\gamma$ is gyromagnetic ratio, $\omega=2\pi f$, $4\pi M_s$ is the saturation magnetization of the ferromagnetic film. The output signal is a result of superposition of all the excited spin waves traveling through the different paths of the matrix. The amplitude of the output spin wave is detected by the voltage generated in the output element (e.g. the inductive voltage produced by the spin waves in the antenna contour). The amplitude of the output voltage is corresponding to the maximum when all the waves are coming in-phase (constructive interference), and the minimum when the waves cancel each other (destructive interference). The output voltage at each port depends on the magnetic states of the nano-magnets within the matrix and the initial phases of the input spin waves. In order to recognize the internal state of the magnonic memory, the initial phases are varied (e.g. from 0 to π). The ensemble of the output values obtained at the different phase combinations constitute a hologram which uniquely corresponds to the internal structure of the matrix.

In general, each of the nanomagnets can have more than 2 thermally stable states, which makes it possible to build a multi-state holographic memory device (i.e. $z^N$ possible memory states, where $z$ is the number of stable magnetic states of a single junction and $N$ is the number of junctions in the magnetic matrix). The practically achievable memory capacity depends on many factors including the operational wavelength, coherence length, the strength of nano-magnets coupling with the spin wave buses, and noise immunity. In the next Section, we present experimental data on the operation of the prototype 2-bit magnonic holographic memory.

### III. EXPERIMENTAL DATA

The set of experiments started with the spin wave transport study in a single cross structure, which is the elementary building block for 2D MHM as depicted in Fig.1. Two types of single cross devices made of $Y_3Fe_2(FeO_4)_3$ (YIG) and Permalloy ($Ni_{81}Fe_{19}$) were fabricated. Both of these materials are promising for application in magnonic waveguides due to their high coherence length of spin waves. At the same time, YIG and Permalloy differ significantly in electrical properties (e.g. YIG is an insulator, permalloy is a conductor) and in fabrication method. YIG cross structures were made from single crystal YIG films epitaxially grown on top of Gadolinium Gallium Garnett ($Gd_3Ga_5O_{12}$) substrates using the liquid-phase transition process. After the films were grown, micro-patterning was done by laser ablation using a pulsed infrared laser (λ≈1.03 μm), with a pulse duration of ~256 ns. The YIG cross junction has the following dimensions: the length of the whole structure is 3mm; the width of the arm in 360μm; thickness is 3.6um. Permalloy crosses were fabricated on top of oxidized silicon wafers. The wafer was spin coated with a 5214E Photoresist at 4000 rpm and exposed using a Karl Suss Mask Aligner. After development, a permalloy metal film was deposited via Electron-Beam Evaporation with a thickness of 100nm and with an intermediate seed layer of 10 nm of Titanium to increase the adhesion properties of the Permalloy film. Lift-off using acetone completed the process. Permalloy cross junction has the following dimensions: the length of the whole structure is 18um; the width of the arm in 6μm; thickness is 100 nm.

Spin waves in YIG and Permalloy structures were excited and detected via micro-antennas that were placed at the edges of the cross arms. Antennas were fabricated from gold wire and mechanically placed directly at the top of the YIG cross. In the case of permalloy, the conducting cross was insulated with a 100nm layer of $SiO_2$ deposited via Plasma-Enhanced Chemical Vapor Deposition (PECVD) and gold antennas were fabricated using the same photolithographic and lift-off procedure as with the permalloy cross structures. A Hewlett-Packard 8720A Vector Network Analyzer (VNA) was used to excite/detect spin waves within the structures using RF frequencies. Spin waves were excited by the magnetic field generated by the AC electric current flowing through the antenna(s). The detection of the transmitted spin waves is via the inductive voltage measurements as described in Ref. [28]. Propagating spin waves change the magnetic flux from the surface, which produces an inductive voltage in the antenna contour. The VNA allowed the S-Parameters of the system to be measured; showing both the amplitude of the signals as well as the phase of both the transmitted and reflected signals. Samples were tested inside a GMW 3472-70 Electromagnet system which allowed the biasing magnetic field to be varied from -1000 Oe to +1000 Oe. The schematics of the experimental setup for spin wave transport study in the single cross structures are shown in Figure 2.

First, we studied spin wave propagation between the four arms of the permalloy cross-structure as shown in Fig.3(A-B) under different bias magnetic field. The input/output ports are numbered from 1 to 4 starting at the 9 O' clock position and then enumerated sequentially in along a clockwise direction. In order to define the angle between the external magnetic field and the direction of signal propagation, we define the X axis along the line from port 1 to port 3, and the Y axis along the line from port 4 to port 2 propagating as depicted in Fig.2. Spin waves were excited on port 2 (the top of the magnetic cross) and read out from port 4 (the bottom of the cross) (see figure 1). The graph in Fig.3 shows the change of the amplitude of the transmitted signal as a function of the strength of the external magnetic field directed perpendicular to the propagating spin waves as depicted in the inset to Fig.3. Hereafter, we show the relative change of the amplitude in decibels normalized to some value (e.g. to the maximum value). The normalization is needed as the input power varies significantly for permalloy and YIG structures as well as for the type of experiment. The reference transmission level is taken at 300 Oe, where the $S_{12}$ parameter is at its absolute maximum. At small magnetic fields below 100 Oe a very small amplitude signal was observed. At approximately 150 Oe there is a noticeable increase in the amplitude followed by a plateau in the response as the field is increased to 500 Oe. Also of interest is the response of the signal as a function of the applied magnetic field direction. In Fig.3(D), we present an example of the experimental data showing the influence of the direction of the bias field on spin waves transport from port 2 to port 4. The results demonstrate prominent change in the amplitude of the transmitted signal [18dB] when the field is applied between 20° and 30°. The main observations of these experiments are the following. (i) Spin wave propagation through the cross junction can be efficiently controlled by the external magnetic field. (ii) Both the amplitude and the direction of the magnetic field can be utilized for spin wave control.

We conducted similar experiments on the YIG single cross device as shown in Figure 4. It was observed that prominent signal modulation could be determined by the direction and the strength of the external magnetic field. In Fig.4, there is shown an example of experimental data on the spin wave transport between ports 2 and 1. The maximum transmission between the orthogonal arms occurs when the field is applied at 68°, while the minimum is seen when the field is applied at 0°. The On/Off ratio for the YIG cross reaches 35dB. Of noticeable interest is also the effect of non-reciprocal spin wave propagation. The two curves in Figure 4(D) show signal propagation from port 2 to port 4, and in the opposite direction from port 4 to port 2. The measurements are done at the same bias magnetic field of 998 Oe. There is a difference of about 5dB for the signals propagating in the opposite direction. The effect is observed in a relatively narrow frequency range (e.g. from 5.2GHz to 5.4GHz). The effect of non-reciprocal spin wave propagation may be of some practical interest for building magnonic diodes, though a more detailed study is required.

Concluding on the spin wave transport in the permalloy and YIG single cross structures, prominent signal modulation has been observed in both cases. For the chosen parameters, the operation frequency is slightly higher for YIG structure (~5GHz) than for permalloy (~3GHz). The speed of signal propagation is slightly faster in permalloy ($3.5 \times 10^6$ cm/s) than in YIG ($3.0 \times 10^6$ cm/s). The difference in the spin wave transport can be attributed to the differences between the intrinsic material properties of YIG and Permalloy as well as the difference in the cross dimensions. It is important to note, that in both cases the level of the power consumption was at the microwatt scale (e.g. 0.1μW-1.0μW for permalloy and 0.5μW-5.0μW for YIG) with no feasible effect of micro heating on the spin wave transport. The summary of the experimental findings for permalloy and YIG single cross junctions cab be found in Table I.

Next, we carried out experiments on spin wave transport and interference in the double-cross structure made of YIG as shown in Figure 5. The choice of material is mainly due to the larger size of the structure and spin wave detecting antennas, where the larger the area of the detecting contour results in higher the observed output inductive voltage. The multi-port double-cross YIG structure is suitable for the study of spin wave interference. In this study, several coherent spin wave signals were excited by ports 2,3,4 and 5 connected to one port of the VNA. The output is detected at port 6. The phase shifters were employed to vary the phase difference between the ports as shown in Figure 5(B). Figure 6 show the experimental data on the output voltage collected in the frequency range from 5.3GHz to 5.5GHz. The curves of the different color correspond to the different phase shifts between the spin wave generated ports. Phase 1 represents a change in the phase of ports 4 and 6 and Phase 2 represents a change in the phase of ports 3 and 5. Figures 6(B-D) show the slices of data taken at a frequency of 5.385GHz, 5.410GHz and 5.45GHz, respectively. The black markers depict the experimentally obtained data, and the red markers depict the theoretical output for the ideal case of the interfering waves of the same frequency and amplitude. The theoretical data is normalized to have the same maximum value as the experimental data at phase difference zero (constructive interference). Taking $l$=1.1mm, $d$=360μm, H=1000 Oe, $4\pi M_s$=1750G, and $\delta H$=20Oe, we estimated possible phase shift by Eqs. (1-2) to be about $\pi/2$, which is in good agreement with the experimental data. This fact implies the dominant role of wave interference in the output signal formation. Discrepancies in the amplitude can be attributed to parasitic noise which raises the base amplitude of the signal to greater than nonzero value even when the phase should be perfectly destructive. Also, it should be noted that Eqs. (1-2) are derived for spin wave propagating in a homogeneous magnetic field, while the magnetic field produced by the micromagnets in the experiment may be inhomogeneous across the thickness and in lateral dimensions.

The data presented in Figure 7 are collected in the experiments where Phase 2 (ports 3 and 5) was changed, while Phase 1 (ports 2 and 4) was kept constant. The ability to independently change the initial phases of the spin waves is equivalent to changing the angle of illumination for building a holograms as illustrated in Fig.1(B). In Figure 7, we present experimental data showing the holographic image of the double-cross structure without memory elements. The surface is a computer reconstructed 3-D plot showing the output voltage as a function of Phase 1 and Phase 2. The excitation frequency is 5.40 GHz, the bias magnetic field is 1000 Oe directed from port 1 toward port 6. In this case, antennas on ports 2 and 4 generated spin waves with the initial Phase 1, and antennas on ports 3 and 5 generated spin waves with initial Phase 2. No signal is applied to port 1. The output is detected at port 6. The change of the output inductive voltage is a result of spin wave interference. It has maximum values in the case of the constructive interference (i.e. Phase 1=Phase 2, (0,0) or ($\pi,\pi$)), and shows minimum output signal when the waves are coming out-of-phase ((0,$\pi$) or ($\pi$,0)).

Finally, we conducted experiments to demonstrate the operation of a prototype 2-bit magnonic holographic memory device. Two micro-magnets made of cobalt magnetic film were placed on top of the junctions of the double-cross YIG structure. As mentioned in Section II, these magnets serve as a memory element, where the magnetic state represents logic zeroes and ones. The schematic of the double-cross structure with micro-magnets attached are shown in Figure 8(A). The length of each magnet is 1.1mm, the width is 360μm and each has a coercivity of 200-500 Oersted (Oe). For the test experiments, we used four mutual orientations of micro-magnets, where the magnets are oriented parallel to the axis connecting ports 1-6, or the axis connecting 2-4; and two cases when the micro-magnets are oriented in the orthogonal directions. Holographic images were collected for each case. Fig.8 shows the collection of data corresponding to output voltage obtained for different magnetic configurations. The phases of the input elements are the same as in the previous experiment. Markers of different shape and color in the legend of figure 8 represent the direction of the "north" end of the micro magnet. The output from the same structure varies significantly for different phase combinations. In some cases, the magnetic states of the magnets can be recognized by just one measurement (e.g. (0,0) phase combination). It is also possible that different magnetic states provide almost the same output (e.g. parallel and orthogonal magnet configurations measured at ($\pi$,0) phase combination). The main observation we want to emphasize is the feasibility of parallel read-out and reconstruction of the magnetic state via spin wave interference. As one can see from the data in Figure 8(B), it is possible to distinctly identify the magnetic states by changing the phases of the interfering waves, which is similar to changing the angle of observation in a conventional optical hologram. We would like to emphasize that all experiments reported in this Section are done at room temperature.

## IV. DISCUSSION

The obtained experimental data show the practical feasibility of utilizing spin waves for building magnonic holographic logic devices and helps to illustrate the advantages and shortcomings of the spin wave approach. Of these results there are several important observations we wish to highlight.

First, spin wave interference patterns produced by multiple interfering waves are recognized for a relatively long distance (more than 3 millimeters between the excitation and detection ports) at room temperature. Despite the initial skepticism [29], coding information into the phase of the spin waves appears to be a robust instrument for information transfer showing a *negligible effect to thermal noise* and immunity to the structures imperfections. This immunity to the thermal fluctuations can be explained by taking into account that the flicker noise level in ferrite structures usually does not exceed -130 dBm[30]. At the same time, spin waves are not sensitive to the structure's imperfections which have dimensions much shorter than the wavelength. These facts explain the good agreement between the experimental and theoretical data (e.g. as shown in Figure 6).

Second, spin wave transport in the magnetic cross junctions is efficiently modulated by an external magnetic field. Spin wave propagation through the cross junction depends on *the amplitude* as well as *the direction* of the external field. This provides a variety of possibilities for building magnetic field-effect logic devices for general and special task data processing. Boolean logic gates such as AND, OR, NOT can be realized in a single cross structure, where an applying external field exceeding some threshold stops/allows spin wave propagation between the selected arms. The ability to modulate spin wave propagation by the direction of the magnetic field is useful for application in non-Boolean logic devices. It is important to note that in all cases the magnitude of the modulating magnetic field is of the order of hundreds of Oersteds, which can be produced by micro- and nano-magnets.

Finally, it appears possible to recognize the magnetic state of the magnet placed on the top of the cross junction via spin waves, which introduces an alternative mechanism for magnetic memory read out. This property itself may be utilized for improving the performance of conventional magnetic memory devices. However, the fundamental advantage of the magnonic holographic memory is the ability to read-out a number of magnetic bits in parallel though the obtained experimental data demonstrates the parallel read-out of just two magnetic bits. In the rest of this Section, we discuss the fundamental limits and the technological challenges of building multi-bit magnonic holographic devices and present the estimates on the device performance.

We start the discussion with the choice of magnetic material for building spin waveguides. Spin wave transport in nanometer scale magnetic waveguides has been intensively studied during the past decade[28, 31-33]. There are two materials that have become predominant, permalloy ($Ni_{81}Fe_{19}$) and YIG, for spin wave devices prototyping. The coherence length of spin waves in permalloy is about tens of microns at room temperature [28, 31], while the coherence length in a non-conducting YIG exceeds millimeters[34]. The attenuation time for spin waves at room temperature is about a nanosecond in permalloy and a hundreds of nanoseconds in YIG[34]. However, the fabrication of YIG waveguides require a special gadolinium gallium garnet (GGG) substrate. In contrast, a permalloy film can be deposited onto a silicon platform by using the sputtering technique. Though YIG has better properties in terms of the coherence length and a lower attenuation, permalloy is more convenient for making magnonic devices on a silicon substrate.

There are two major physical mechanisms affecting the amplitude/phase of the spin wave propagating under the junction magnet: (i) interaction with magnetic field produced by the magnet, and (ii) damping due to the presence of the conducting material. The effect of conducting films on spin wave propagation has been studied for MSSWs in the ferrite-metal structures [35, 36]. It was found that the strength of the spin wave dispersion modification is defined by the critical parameter G given as following:

$$G = \frac{t}{q \cdot l_{sk}^2} \quad (3)$$

where $t$ is the thickness of the conducting film, $q$ is the wave number, and $l_{sk}$ is the skin depth . The presence of a metallic film results in a prominent spin wave dispersion modification for G>3, if the is width of the gap between the ferrite and metallic film is less than the wavelength. Spin wave is completely damped in the range 1/3<G<3 due to the excessive absorption by the conducting electrons. The effect vanishes for G<1/3. In our experiments, we used junction magnets made of cobalt with the thickness of 50nm (bulk electric conductivity σ≈1.6×10$^7$ S/m), which correspond to $l_{sk}$ ≈1.5 µm. The range of wave numbers $q$ is restricted by the size of the sample L q>π/L>10 cm$^{-1}$ , and by the width of the micro-antennas W≈30µm, q<π/W<1000 cm$^{-1}$. In all experiments, the range of the wave numbers is confined within the following range: 10 cm$^{-1}$<q<1000 cm$^{-1}$, which, in turn, corresponds to the range for parameter G: 0.002<G<2. It should be noted that Eq.3 has been derived for an infinite ferrite waveguide, which ignores the effect of space confinement. Taking into account the real dimensions of the YIG substrate 3.6µm, we obtain the minimum boundary for q≈80 cm$^{-1}$. Thus, we have G<1/3 (negligible effect of spin wave dispersion modification due to the losses) for all experiments. In general, there may be other physical phenomena contributing to the spin wave dispersion modification in a magnetic cross-structure. The development of MHM devices will require a great deal of efforts in the theoretical study and numerical modeling of spin wave transport in magnetic nanostructures.

In order to make a multi-bit magnonic holographic devices, the operating wavelength should be scaled down below 100nm [23]. The main challenge with shortening the operating wavelength is associated with the building of nanometer-scale spin wave generating/detecting elements. There are several possible ways of building input/output elements by using micro-antennas[28], spin torque oscillators[26], and multi-ferroic elements [6]. So far, micro-antennas are the most convenient and widely used tool for spin wave excitation and detection in ferromagnetic films[31]. Reducing the size of the antenna will lead to the reduction of the detected inductive voltage. This fact limits the practical application of any types of conducting contours for spin wave detection. The utilization of spin torque oscillators makes it possible to scale down the size of the elementary input/output port to several nanometers [37]. The main challenge for the spin torque oscillators approach is to reduce the current required for spin wave generation. More energetically efficient are the two-phase composite multiferroics comprising piezoelectric and magnetostrictive materials [38]. An electric field applied across the piezoelectric produces stress, which, in turn, affects the magnetization of the magnetoelastic material. The advantage of the multiferroic approach is that the magnetic field required for spin wave excitation is produced via magneto-electric coupling by applying an alternating electric field rather than an electric current. For example, in Ni/PMN-PT synthetic multiferroic reported in Ref. [39], an electric field of 0.6MV/m has to be applied across the PMN-PT in order to produce 90 degree magnetization rotation in Nickel. Such a relatively low electric field required for magnetization rotation translates in ultra-low power consumption for spin wave excitation [20]. At the same time, the dynamics of the synthetic multiferroics, especially at the nanometer scale, remains mostly unexplored.

To benchmark the performance of the magnonic holographic devices, we apply the charge-resistance approach as developed in Ref. [40] The details of the estimates and the key assumptions are given in Appendix A. According to the estimates, MHM device consisting of 32 inputs, with a 60nm separation distance between the inputs would consume as low as 150µW of power or 72fJ per computation. At the same time, the functional throughput of the MHM scales proportional to the number of cells per area/volume and exceeds $1.5 \times 10^{18}$ bits/cm$^2$/s for a 60nm feature size. It is interesting to note, that holographic logic units can be used for solving certain nondeterministic polynomial time (NP) class of problems (i.e. finding the period of the given function). The efficiency of holographic computing with classical waves is somewhere intermediate between digital logic and quantum computing, allowing us to solve a certain class of problems fundamentally more efficient than general-type processors but without the need for quantum entanglement[41]. Image recognition and processing are among the most promising applications of magnonic holographic device exploiting its ability to process a large number of bits/pixels in parallel within a single core.

There are many questions on spin wave transport (e.g. in magnetic crosses), which remain mostly unexplored. For instance, it is not clear the mechanism responsible for spin wave splitting between the orthogonal arms. To the best of our knowledge, there is no theoretical work explaining the observed spin wave propagation in magnetic crosses. It would be of great interest to study the dynamics of spin wave redirection depending on the geometry of the cross. It may be expected that the amplitude/phase of the redirected (bended) spin wave depends on the wavelength/size ratio, the material properties of the cross, and magnetic field produced by the nano-magnet. Also, in this work, we attribute the change in the interference pattern to the different phase shifts accumulated by spin waves propagating under the nano-magnets of different orientation (i.e. Eqs. 1-2). A real picture may be much more complicated due to the difference in amplitudes, which may arise to the different factors. There is no doubt that the development of magnonic holographic memory devices will take a great deal of efforts including experimental as well as theoretical studies.

## V. CONCLUSIONS

The collected experimental data show rich physical phenomena associated with spin wave propagation in single- and double-cross structures. Prominent signal modulation by the direction, rather than the amplitude of magnetic field and the low effect of thermal noise on spin wave propagation at room temperature are among the many interesting findings presented here. The effect of spin wave redirection between the cross arms by the external magnetic field may be further exploited for building a variety of logic devices. Besides, spin waves appear to be a robust instrument allowing us to sense the magnetic state of micro-magnets by the change in the interference pattern. Quite surprisingly, it is possible to recognize the unique holographic output for the different orientations of micro-magnets in a relatively long device at room temperature. Overall, the obtained data demonstrates the practical feasibility of building magnonic holographic devices. These holographic devices are aimed not to replace but to complement CMOS in special type data processing such as speech recognition and image processing. According to estimates, scaled magnonic holographic devices may provide more than $1 \times 10^{18}$ bits/cm$^2$/s data processing rate while consuming less than 0.2mW of energy. The main technological challenges are associated with the scaling down the operating wavelength and building nanometer scale spin wave generating/detecting elements with spin torque oscillators and multiferroics being among the most promising solutions. At the same time, it is expected that reducing the operating wavelength will make spin waves more sensitive to structure imperfections. The development of scalable magnonic holographic devices and their incorporation with conventional electronic devices may pave the road to the next generations of logic devices with functional capabilities far beyond current CMOS.

APPENDIX

This section contains the details on the power consumption estimates. We assume that the input to each magnetoelectric (ME) cell is an AC voltage with amplitude $V_{in}$ created in a RLC oscillator. Then the power dissipation on resonance is

$$P_{in} = \frac{V_{in}^2}{2R}.$$

The amplitude of the electric field in the piezoelectric of thickness $t_{pz}$ is

$$E_{in} = V_{in} / t_{pz}.$$

The amplitude of strain created in the piezoelectric of the ME cell is

$$\varepsilon_{xx} = d_{31} E_{in},$$

where the piezoelectric coefficient is $d_{31}$. Hereafter, it is assumed that spin wave is propagating along the X axis as shown in the inset to Fig2. The stress transferred to the ferromagnet is

$$\sigma = Y \varepsilon_{xx},$$

where the Young's modulus of the piezoelectric is $Y$. The change in the magnetic anisotropy due to magnetostriction is

$$U_{ms} = \frac{3}{2} \lambda \sigma,$$

where the magnetostriction coefficient of the ferromagnet is $\lambda$. Then the maximum amplitude of magnetization change is

$$\Delta M_x \approx \frac{2 U_{ms}}{\mu_0 M_s},$$

where the permeability of vacuum is , and the saturation magnetization is $M_s$. This can be expressed via the magnetoelectric coefficient

$$\alpha \approx \frac{\mu_0 \Delta M_x}{E_{in}} \approx \frac{3 \lambda Y d_{31}}{M_s}.$$

Then the generated dimensionless amplitude of the spin wave can be approximated as follows

$$A_{in} = \frac{\Delta M_x}{M_s} \approx \frac{\alpha V_{in}}{\mu_0 M_s t_{pz}},$$

The spin waves interact and attenuate as they propagate. If the distance between inputs is $L$, the number of inputs is $N$, and the attenuation length is $l_{at}$, then the propagation distance is

$$L_{tot} = NL \tag{9}$$

and let the minimum amplitude needed to be detected is a quarter of the average output amplitude

$$A_{min} = \frac{A_{in}}{4} \exp\left(-\frac{L_{tot}}{l_{at}}\right). \tag{10}$$

If the group velocity of the spin waves is $\tau = L_{tot}/c_{sw}$, then the time needed for one holographic imaging is

$$\tau = NL/c_{sw}. \tag{11}$$

Assuming that the detection occurs by the inverse of the ME effect, and that its coefficients are the same as for the direct ME effect, we obtain that the connection between the magnetization amplitude and the amplitude of the generated electric field

$$\alpha \approx \frac{\varepsilon_{pz}\varepsilon_0 E_{min}}{\Delta M}. \tag{12}$$

Thus the minimum output voltage is

$$V_{min} = E_{min} t_{pz} = \frac{\alpha M_s A_{min} t_{pz}}{\varepsilon_{pz}\varepsilon_0}. \tag{13}$$

Combining expressions (10) and (13), we have the following for voltages

$$V_{min} = \frac{V_{in}}{4} \exp\left(-\frac{NL}{l_{at}}\right) \frac{\alpha^2}{\varepsilon_{pz}\varepsilon_0\mu_0}. \tag{14}$$

Then the total driving power for $N$ inputs is

$$P_{tot} = \frac{NV_{in}^2}{2R} = \frac{8NV_{min}^2}{R} \exp\left(\frac{2NL}{l_{at}}\right) \frac{\varepsilon_{pz}^2}{c^4\alpha^4}, \tag{15}$$

where $c$ is the speed of light. For minimum detectability, the output voltage should exceed the Johnson noise voltage by 5X. The spectral density of noise is

$$V_n^2 = 4k_B TR. \tag{16}$$

The required power within the bandwidth $B$ (approximately equal to the ac voltage frequency) is

$$P_{tot} = 800NBk_B T \exp\left(\frac{2NL}{l_{at}}\right) \frac{\varepsilon_{pz}^2}{c^4\alpha^4}. \tag{17}$$

And the total energy for one imaging is

$$E_{tot} = P_{tot}\tau. \tag{18}$$

Using the magnetostriction parameters for the most advantageous case of Terfenol-D and PMN-PT, the magnetoelectric coefficient

$$\alpha = 57 ns/m \approx 17/c.$$

The dielectric constant of PMN-PT is now $\varepsilon_{pz} = 1000$.

Substituting the parameters of the holographic system:
$N = 32$, $L = 60 nm$, $c_{sw} = 4000 m/s$, $B = 100 GHz$, $l_{at} = 24 \mu m$, we obtain:
$\tau = 480 ps$, $P_{tot} = 150 \mu W$, $E_{tot} = 72 fJ$.

ACKNOWLEDGMENT

This work was supported in part by the FAME Center, one of six centers of STARnet, a Semiconductor Research Corporation program sponsored by MARCO and DARPA and by the National Science Foundation under the NEB2020 Grant ECCS-1124714.

REFERENCES

[1] I. T. R. Semiconductors, "2011," http://www.itrs.net, vol. Chapter PIDS.
[2] K. Bernstein, R. K. Cavin, W. Porod, A. Seabaugh, and J. Welser, "Device and Architecture Outlook for Beyond CMOS Switches," *Proceedings of the IEEE*, vol. 98, pp. 2169-84, 2010.
[3] D. E. Nikonov and I. A. Young, "Overview of Beyond-CMOS Devices and a Uniform Methodology for Their Benchmarking," *Proceedings of the IEEE*, vol. 101, pp. 2498-2533, Dec 2013.
[4] G. E. Moore, "Gramming more components onto integrated circuits," *Electronics*, vol. 38, pp. 114-117, 1965.
[5] G. Bourianoff, J. E. Brewer, R. Cavin, J. A. Hutchby, and V. Zhirnov, "Boolean Logic and Alternative Information-Processing Devices," *Computer*, vol. 41, pp. pp. 38-46, 2008
[6] S. Cherepov, P. Khalili, J. G. Alzate, K. Wong, M. Lewis, P. Upadhyaya, J. Nath, M. Bao, A. Bur, T. Wu, G. P. Carman, A. Khitun, and K. L. Wang, "Electric-field-induced spin wave generation using multiferroic magnetoelectric cells," *Proceedings of the 56th Conference on Magnetism and Magnetic Materials (MMM 2011), DB-03, Scottsdale, Arizona* 2011.
[7] K. Perzlmaier, M. Buess, C. H. Back, V. E. Demidov, B. Hillebrands, and S. O. Demokritov, "Spin-wave eigenmodes of permalloy squares with a closure domain structure," *Physical Review Letters*, vol. 94, Feb 11 2005.
[8] S.-K. Kim, K.-S. Lee, and D.-S. Han, "A gigahertz-range spin-wave filter composed of width-modulated nanostrip magnonic-crystal waveguides," *Applied Physics Letters*, vol. 95, Aug 24 2009.
[9] Y. Au, M. Dvornik, O. Dmytriiev, and V. V. Kruglyak, "Nanoscale spin wave valve and phase shifter," *Applied Physics Letters*, vol. 100, Apr 23 2012.
[10] G. Gubbiotti, M. Conti, G. Carlotti, P. Candeloro, E. Di Fabrizio, K. Y. Guslienko, A. Andre, C. Bayer, and A. N. Slavin, "Magnetic field dependence of quantized and localized spin wave modes in thin rectangular magnetic dots," *Journal of Physics-Condensed Matter*, vol. 16, pp. 7709-7721, Nov 3 2004.
[11] S. Mansfeld, J. Topp, K. Martens, J. N. Toedt, W. Hansen, D. Heitmann, and S. Mendach, "Spin Wave Diffraction and Perfect Imaging of a Grating," *Physical Review Letters*, vol. 108, Jan 26 2012.
[12] K. Perzlmaier, G. Woltersdorf, and C. H. Back, "Observation of the propagation and interference of spin waves in ferromagnetic thin films," *PHYSICAL REVIEW B*, vol. 77, Feb 2008.
[13] V. E. Demidov, S. O. Demokritov, K. Rott, P. Krzysteczko, and G. Reiss, "Mode interference and periodic self-focusing of spin waves in permalloy microstripes," *PHYSICAL REVIEW B*, vol. 77, Feb 2008.
[14] S. V. Vasiliev, V. V. Kruglyak, M. L. Sokolovskii, and A. N. Kuchko, "Spin wave interferometer employing a local nonuniformity of the effective magnetic field," *JOURNAL OF APPLIED PHYSICS*, vol. 101, Jun 1 2007.
[15] M. Jamali, J. H. Kwon, S.-M. Seo, K.-J. Lee, and H. Yang, "Spin wave nonreciprocity for logic device applications," *Scientific Reports*, vol. 3, Nov 7 2013.
[16] M. Hamali, J. Kwon, S. Seo, K. Lee, and H. Yang, "Spin wave nonreciprocity for logic device applications," *Scientific Reports*, 2013.

[17]  M. P. Kostylev, A. A. Serga, T. Schneider, B. Leven, and B. Hillebrands, "Spin-wave logical gates," *Applied Physics Letters,* vol. 87, pp. 153501-1-3, 2005.
[18]  T. Schneider, A. A. Serga, B. Leven, B. Hillebrands, R. L. Stamps, and M. P. Kostylev, "Realization of spin-wave logic gates," *Appl. Phys. Lett.,* vol. 92, pp. 022505-3, 2008.
[19]  L. Ki-Suk and K. Sang-Koog, "Conceptual design of spin wave logic gates based on a Mach-Zehnder-type spin wave interferometer for universal logic functions," *JOURNAL OF APPLIED PHYSICS,* vol. 104, pp. 053909 (4 pp.)-053909 (4 pp.), 1 Sept. 2008.
[20]  A. Khitun and K. L. Wang, "Non-volatile magnonic logic circuits engineering," *JOURNAL OF APPLIED PHYSICS,* vol. 110, Aug 1 2011.
[21]  Y. Wu, M. Bao, A. Khitun, J.-Y. Kim, A. Hong, and K. L. Wang, "A Three-Terminal Spin-Wave Device for Logic Applications," *Journal of Nanoelectronics and Optoelectronics,* vol. 4, pp. 394-397, Dec 2009.
[22]  P. Shabadi, A. Khitun, P. Narayanan, B. Mingqiang, I. Koren, K. L. Wang, and C. A. Moritz, "Towards logic functions as the device," *2010 IEEE/ACM International Symposium on Nanoscale Architectures (NANOARCH 2010),* 01 2010.
[23]  A. Khitun, "Magnonic holographic devices for special type data processing," *JOURNAL OF APPLIED PHYSICS,* vol. 113, Apr 28 2013.
[24]  M. Covington, T. M. Crawford, and G. J. Parker, "Time-resolved measurement of propagating spin waves in ferromagnetic thin films (vol 89, art no 237202, 2002)," *Physical Review Letters,* vol. 92, Feb 27 2004.
[25]  T. J. Silva, C. S. Lee, T. M. Crawford, and C. T. Rogers, "Inductive measurement of ultrafast magnetization dynamics in thin-film Permalloy," *JOURNAL OF APPLIED PHYSICS,* vol. 85, pp. 7849-7862, Jun 1 1999.
[26]  S. Kaka, M. R. Pufall, W. H. Rippard, T. J. Silva, S. E. Russek, and J. A. Katine, "Mutual phase-locking of microwave spin torque nano-oscillators," *Nature,* vol. 437, pp. 389-392, 2005.
[27]  J. Eschbach and R. Damon, *J. Phys. Chem. Solids,* vol. 19, p. 308, 1961.
[28]  M. Covington, T. M. Crawford, and G. J. Parker, "Time-resolved measurement of propagating spin waves in ferromagnetic thin films," *Physical Review Letters,* vol. 89, pp. 237202-1-4, 2002.
[29]  S. Bandyopadhyay and M. Cahay, "Electron spin for classical information processing: a brief survey of spin-based logic devices, gates and circuits," *Nanotechnology,* vol. 20, Oct 14 2009.
[30]  E. Rubiola, Y. Gruson, and V. Giordano, "On the flicker noise of ferrite circulators for ultra-stable oscillators," *Ieee Transactions on Ultrasonics Ferroelectrics and Frequency Control,* vol. 51, pp. 957-963, Aug 2004.
[31]  T. J. Silva, C. S. Lee, T. M. Crawford, and C. T. Rogers, "Inductive measurement of ultrafast magnetization dynamics in thin-film Permalloy," *Journal of Applied Physics,* vol. 85, pp. 7849-62, 1999.
[32]  M. Bailleul, D. Olligs, C. Fermon, and S. Demokritov, "Spin waves propagation and confinement in conducting films at the micrometer scale," *Europhysics Letters,* vol. 56, pp. 741-7, 2001.
[33]  V. E. Demidov, J. Jersch, K. Rott, P. Krzysteczko, G. Reiss, and S. O. Demokritov, "Nonlinear Propagation of Spin Waves in Microscopic Magnetic Stripes," *Physical Review Letters,* vol. 102, p. 177207, 2009.
[34]  A. A. Serga, A. V. Chumak, and B. Hillebrands, "YIG magnonics," *Journal of Physics D-Applied Physics,* vol. 43, Jul 7 2010.
[35]  A. G. Veselov, S. L. Vysotsky, G. T. Kazakov, A. G. Sukharev, and Y. A. Filimonov, "Surface magnetostatic waves in metallized YIG films Journal of communication technologies and electronics," *Journal of communication technologies and electronics,* vol. 39, pp. 2067-2074, 1994.
[36]  Y. A. Filimonov and Y. V. Khivintsev, "The interaction of surface magnetostatic and bulk elastic waves in metallized ferromagnet-insulator structure," *Journal of communication technologies and electronics,* vol. 47, pp. 1002-1007, 2002.
[37]  M. Madami, S. Bonetti, G. Consolo, S. Tacchi, G. Carlotti, G. Gubbiotti, F. B. Mancoff, M. A. Yar, and J. Akerman, "Direct observation of a propagating spin wave induced by spin-transfer torque," *Nature Nanotechnology,* vol. 6, pp. 635-638, Oct 2011.
[38]  K. Roy, S. Bandyopadhyay, and J. Atulasimha, "Energy dissipation and switching delay in stress-induced switching of multiferroic nanomagnets in the presence of thermal fluctuations," *JOURNAL OF APPLIED PHYSICS,* vol. 112, Jul 15 2012.
[39]  T. Wu, A. Bur, P. Zhao, K. P. Mohanchandra, K. Wong, K. L. Wang, C. S. Lynch, and G. P. Carman, "Giant electric-field-induced reversible and permanent magnetization reorientation on magnetoelectric Ni/(011) [Pb(Mg1/3Nb2/3)O3](1−x)–[PbTiO3]x heterostructure," *Applied Physics Letters,* vol. 98, pp. 012504-7, 2011.
[40]  A. Sarkar, D. E. Nikonov, I. A. Young, B. Behin-Aein, and S. Datta, "Charge-resistance approach to benchmarking performance of beyond-CMOS information processing devices," *Ieee Transactions on Nanotechnology,* vol. 13, pp. 143-50, Jan. 2014.
[41]  L. G. Valiant, "Holographic algorithms," *Proceedings. 45th Annual IEEE Symposium on Foundations of Computer Science,* pp. 306-15, 2004 2004.

Figure Captions

Figure 1. (A) Schematics of Magnonic Holographic Memory consisting of a 4×4 magnetic matrix and an array of spin wave generating/detecting elements.  For simplicity, the matrix is depicted as a two-dimensional grid of magnetic wires with just 4 elements on each side. These wires serve as a media for spin wave propagation. The nano-magnet on the top of the junction is a memory element, where information is encoded into the magnetization state. The spins of the nano-magnet are coupled to the spins of the magnetic wires via the dipole-dipole or exchange interaction. (B) Illustration of the principle of operation. Spin waves are excited by the elements on one or several sides of the matrix (e.g. left side), propagate through the matrix and detected on the other side (e.g. right side) of the structure. All input waves are of the same amplitude and frequency. The initial phases of the input waves are controlled by the generating elements. The output waves are the results of the spin wave interference within the matrix.  The amplitude of the output wave depends on the initial phases and the magnetic states of the junctions.

Figure 2. Schematics of the experimental setup for single cross structures testing.  The input and the output micro-antennas are connected to the Hewlett-Packard 8720A Vector Network Analyzer (VNA). The VNA generates input RF signal and measures the S parameters showing the amplitude and the phases of the transmitted and reflected signals. The device under study is placed inside a GMW 3472-70 Electromagnet system which allows the biasing magnetic field to be varied from -1000 Oe to +1000 Oe.  The in-plane axes X and Y are defined along the lines from port 1 to port 3, and from port 4 to port 2, respectively.

Figure 3. (A) Microscope image of Permalloy single cross structure with antennas placed on the edges of the structure. The length of the whole structure is 18um; the width of the arm in 6μm; thickness is 40 nm. (B) Photo of the packaged device with microwave input/output ports used for connection to VNA.  (C) Experimental data showing the relative change of the output signal (inductive voltage) as a function of the strength of the external magnetic field. The output is normalized to the maximum output detected at 300 Oe.  The signal is transmitted from port 2 to port 4, the bias magnetic field is along the X axis as depicted in the inset. The input frequency is 3.16GHz. (D)  Experimental data showing the relative change of the output signal amplitude as a function of the direction of the external magnetic field. The output is normalized to the maximum amplitude at zero degrees (parallel to X axis). The signal is transmitted from port 2 to port 4. The measurements are taken at the different angles $\alpha$ of the bias magnetic field of 148Oe, where $\alpha$ is defined as the angle to the X axis as depicted in the inset.

Figure 4. Microscope image of YIG single cross structure. The length of the whole structure is 3mm; the width of the arm in 360μm; thickness is 3.6um. (B) Experimental data showing the relative change of the output signal amplitude as a function of the direction of the external magnetic field. The signal is transmitted from port 2 to port 1. (C) Experimental data on the signal transmission from port 4 to port 2 (black curve) and from port 2 to port 4 (red curve). The data show about 5dB difference for the signal propagating in the opposite directions in the frequency range from 5.2GHz to 5.5GHz.

Figure 5. (A) Microscope image of YIG double cross structure. The length of the whole structure is 3mm; the width of the arm in 360μm; thickness is 3.6um. (B) Schematics of the double-cross device with six micro-antennas fabricated on the edges under study. (C) Schematics of the experimental setup. The input and the output micro-antennas are connected to the Hewlett-Packard 8720A Vector Network Analyzer (VNA). There is asset of splitters (depicted as S), attenuators (depicted as A), and phase shifters (depicted as Ph) used for the connections with VNA. The device under study is placed inside a GMW 3472-70 Electromagnet system.

Figure 6. Experimental data on the output voltage at port 1 as a result of spin wave interference. The data are collected in the frequency range from 5.3GHz to 5.5GHz. The bias magnetic field is 1000 Oe directed from port 1 toward port 6. The curves of the different color correspond to the different phase shifts between the spin wave generated ports. Phase 1 represents a change in the phase of ports 4 and 6 and Phase 2 represents a change in the phase of ports 3 and 5. (B-D) Slices of data taken at the frequencies of 5.385GHz, 5.410GHz and 5.45GHz, respectively. The black markers depict the experimentally obtained data, and the red markers depict the theoretical data for the ideal case of the interfering waves of the same frequency and amplitude. The theoretical data is normalized to have the same maximum value as the experimental data at phase difference zero (constructive interference).

Figure 7. Holographic image of the double-cross structure without memory elements. The cyan surface is a computer reconstructed 3-D plot based on the experimental data: output voltage as a function of the phases of the interfering spin waves. The output is detected at port 6. The excitation frequency is 5.40 GHz, the bias magnetic field is 1000 Oe directed from port 1 toward port 6. No signal is applied to port 1. The legend and schematics on the right side explain the phases of the spin waves generated at the six ports. Phase 1 and Phase 2 correspond to the phases generated at the ports 2,4 and 3,5, respectively.

Figure 8. Collection of experimental data showing the output of the double-cross structure with micro-magnets placed on the top of the junctions. The phase coordinates show the combination of the initial phases of the spin waves, where Phase 1 and Phase 2 are defined the same way as in Fig.7 (i.e. 0,π)means that the spin waves generated in the ports 2,4 and 3,5 have a-π difference in the initial phase. The markers of different shape and color correspond to the different magnetic configurations as illustrated on the right side. All results are obtained at room temperature.

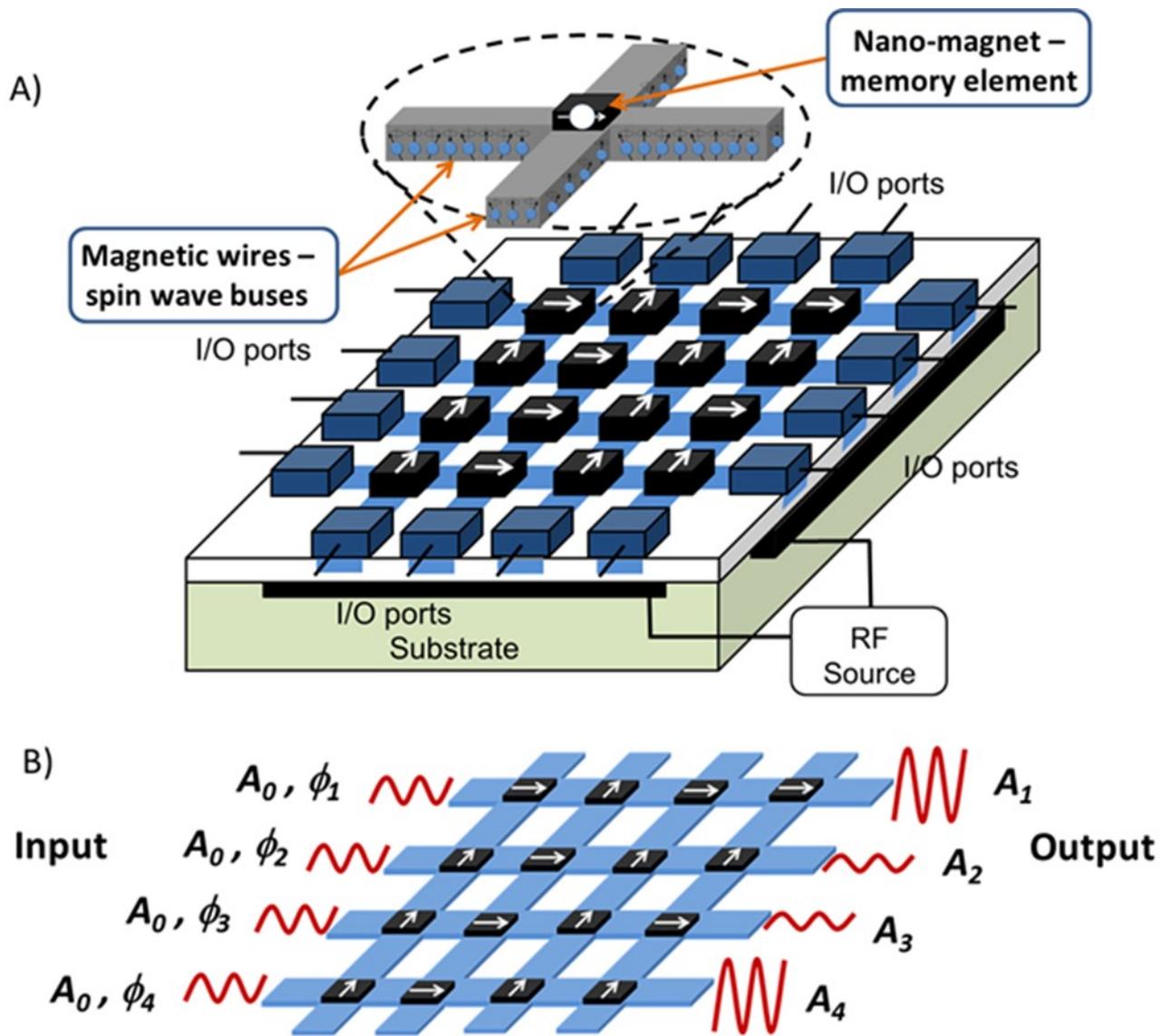

**Figure 1**

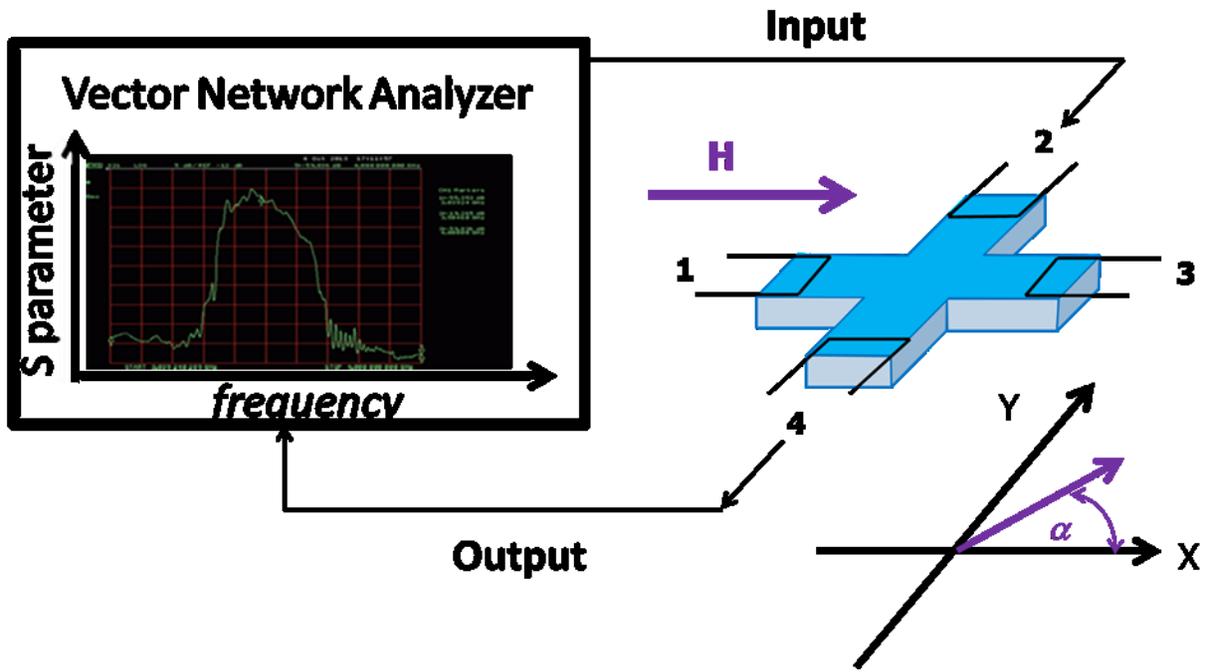

**Figure 2**

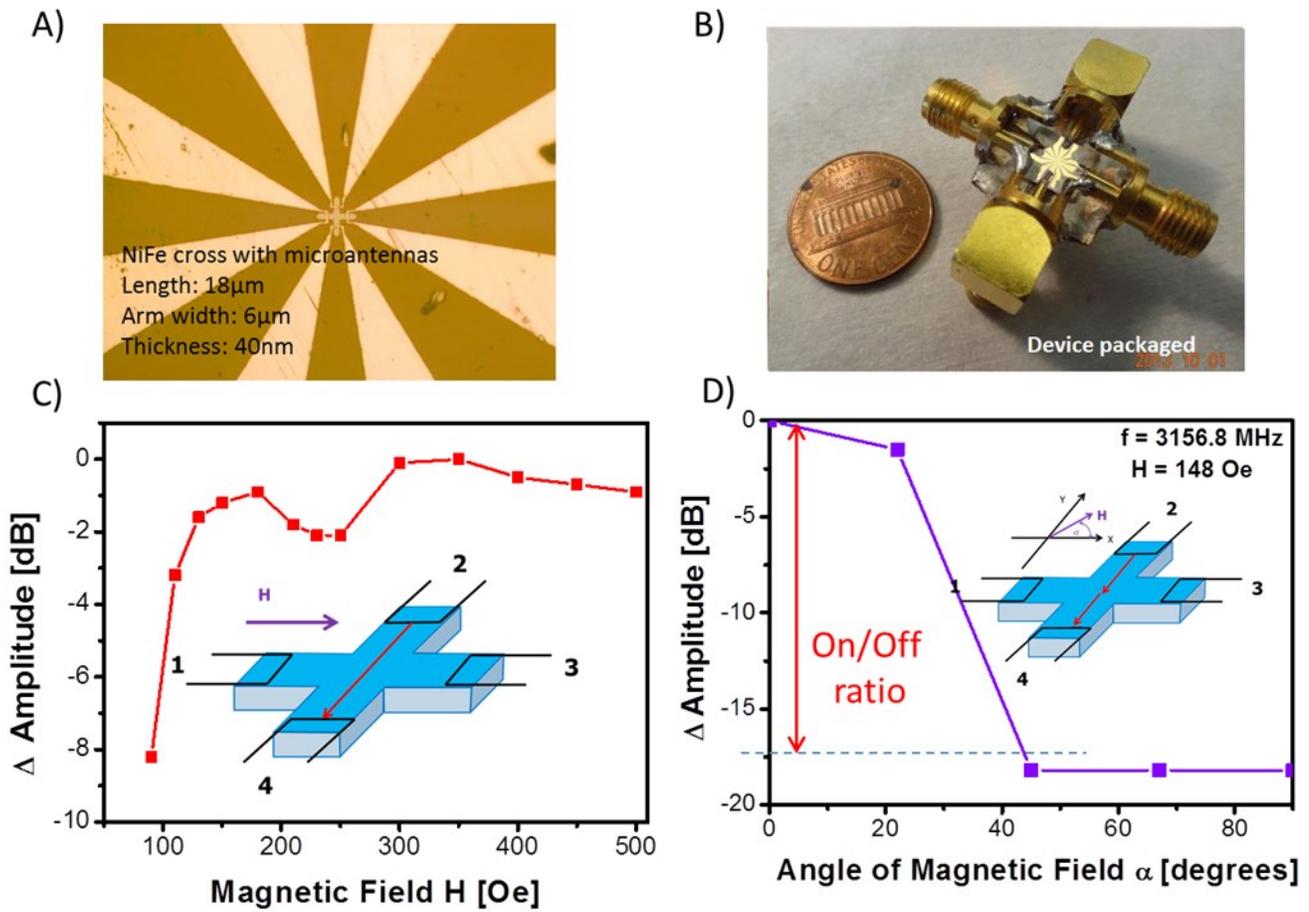

Figure 1

Figure 3

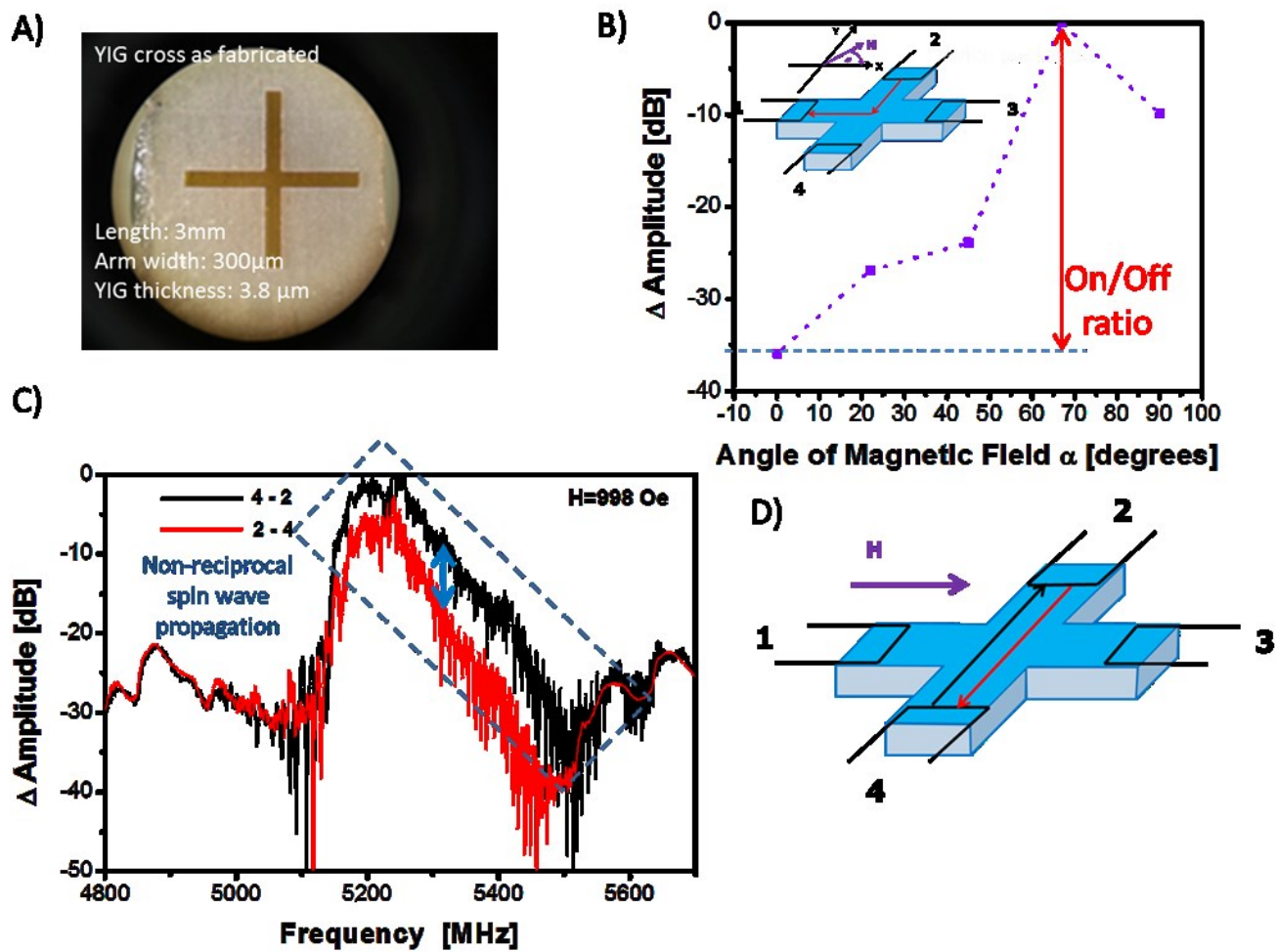

**Figure 4**

**Table I. Summary on the experimental data collected for permalloy and YIG single cross structures**

|  | Permalloy | YIG |
|---|---|---|
| **Cross dimensions** | L=18µm, w=6µm, d=100nm | L=3mm, w=300µm, d=3.8µm |
| **Operational Frequency** | 3GHz-4GHz | 5GHz-6GHz |
| **SW group velocity** | $3.5 \times 10^6$ cm/s | $3.0 \times 10^6$ cm/s |
| **Maximum On/Off ratio** | 20dB | 35dB |
| **Power consumption** | 0.1µW-1µW | 0.5µW-5µW |
| **Compatibility with Silicon** | Yes | No |

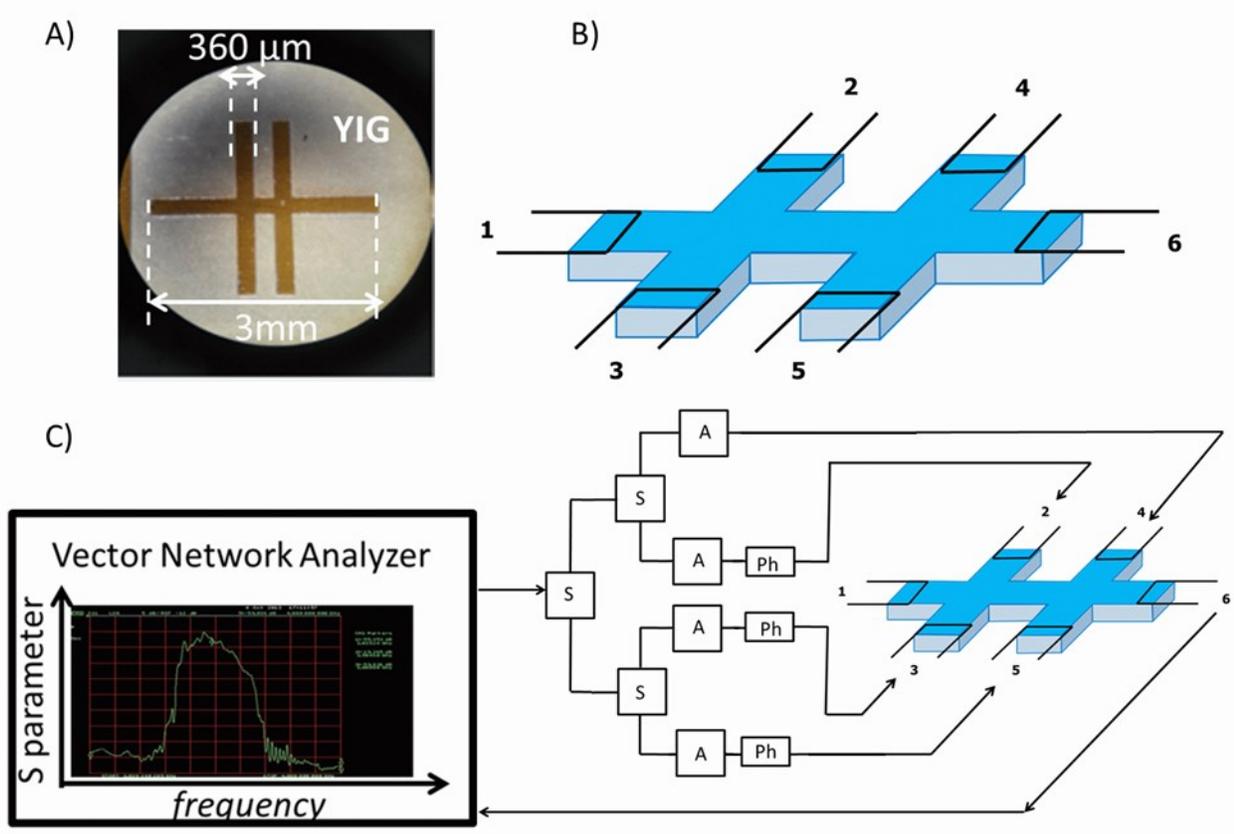

**Figure 5**

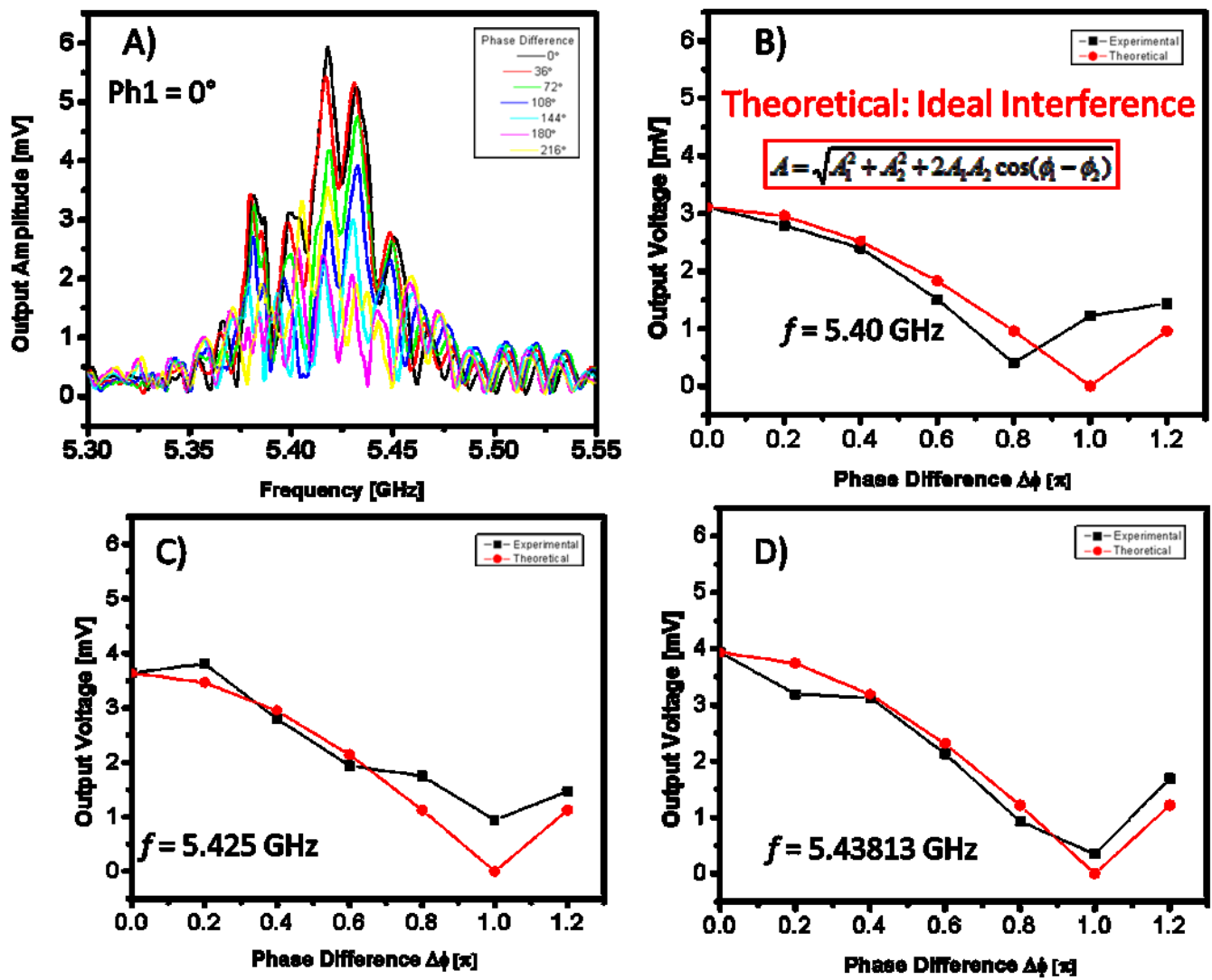

**Figure 6**

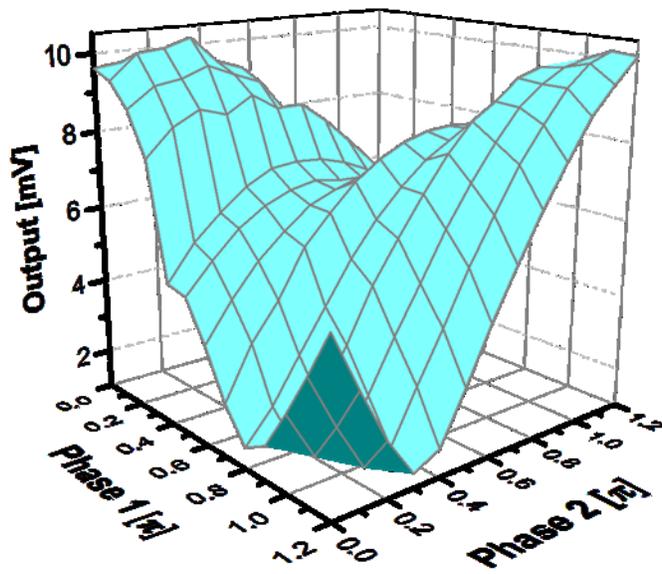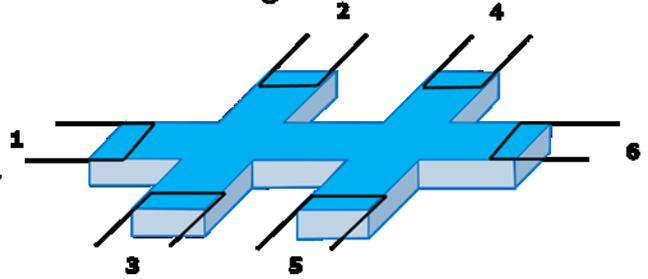

**Figure 7**

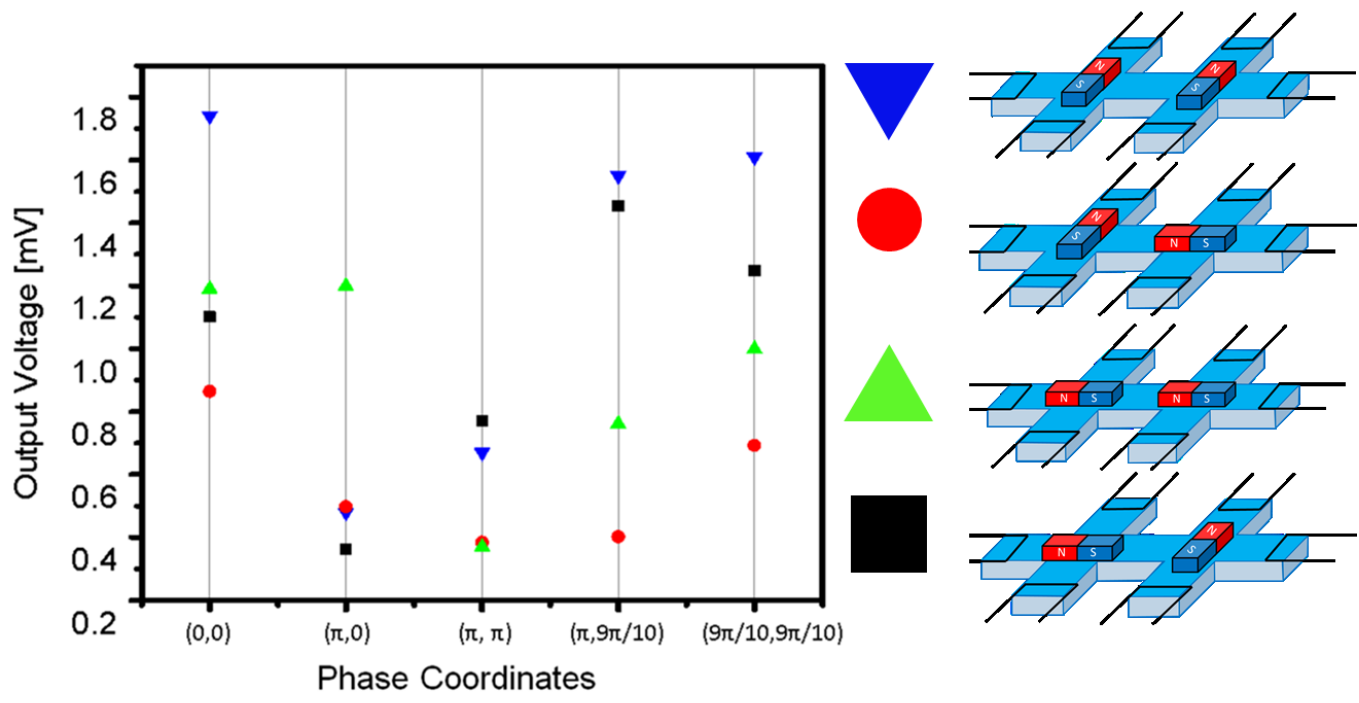

**Figure 8**